# QUAVER: Quantum Unfoldment through Visual Engagement and Storytelling Resources

Ishan Shivansh Bangroo[1], Samia Amir[2]

*University Of Florida*[1]

*University of Haripur*[2]

A B S T R A C T

The task of providing effective instruction and facilitating comprehension of resources is a substantial difficulty in the field of Quantum Computing, mostly attributable to the complicated nature of the subject matter. Our research-based observational study "*QUAVER*" is rooted on the premise that the use of visual tools and narrative constructions has the potential to significantly augment comprehension and involvement within this domain. Prominent analytical techniques, such as the two-sample t-test, revealed a significant statistical difference between the two groups, as shown by the t-statistic and p-value, highlighting the considerable effectiveness of the visual-narrative strategy. One crucial aspect of our study is on the implementation of an exciting algorithmic framework designed specifically to optimize the integration of visual and narrative components in an integrated way. This algorithm utilizes sophisticated heuristic techniques to seamlessly integrate visual data and stories, offering learners a coherent and engaging instructional experience. The design of the material effectively manages the interplay between visual signals and narrative constructions, resulting in an ideal level of engagement and understanding for quantum computing subject. The results of our study strongly support the alternative hypothesis, providing evidence that the combination of visual information and stories has a considerable positive impact on participation in quantum computing education. This study not only introduces a significant approach to teaching quantum computing but also demonstrates the wider effectiveness of visual and narrative aids in complex scientific education in the digital age.

## 1. Introduction

The field of quantum education, which is a crucial frontier in modern science, has always been recognized as an area of significant intricacy and depth. The fundamental principles of quantum physics, as expounded by Bell et al., provide a profound challenge to the conventional understanding of classical intuition. Nevertheless, Bohr contended that those who do not experience astonishment upon encountering quantum theory have not really comprehended it. Nonetheless, the present-day digital era offers a chance to transform this first shock into a state of profound comprehension.

The field of quantum physics, characterized by its inherent abstraction and counter-intuitive nature, has been a significant obstacle for both educators and learners. According to the eminent physicist Feynman, it can be confidently said that there is a general lack of comprehension about the principles of quantum mechanics. The attitude expressed aligns with Bell's examination of the difficulties posed by quantum entanglement for classical intuitions.

However, in the current age of digital enlightenment, one must consider how modern teaching tools and methodologies might effectively address the issue of bridging the gap in quantum cognition. The concept of using visual aids in the field of education is not an unfamiliar one. Paivio previously established the notion of dual coding, positing that textual and visual information are processed via distinct cognitive pathways. Mayer's following studies further solidified the effectiveness of this strategy, highlighting the significant impact of visual aids on the enhancement of learning.

Narrative constructions, as well, are not a recent innovation within the realm of education. According to Bruner, narrative modes of representation are crucial in the process of generating reality and comprehending the universe. In accordance with Haven's claim on the brain's inclination towards narratives, it becomes evident that incorporating quantum physics into storytelling might potentially facilitate a greater level of comprehension.

Notwithstanding the complexities inherent in quantum physics, the fundamental aspect of any instructional undertaking remains firmly rooted in the realm of proficient communication. Throughout history, visual representations and narrative patterns have shown to be advantageous for conveying complicated concepts across several domains. These tools serve as cognitive bridges, aiding the understanding of intricate and abstract ideas by making them more accessible and intelligible. Numerous teaching techniques have flourished based on this fundamental premise, resulting in a significant transformation of the educational environment.

In the current era of digital information, educators have access to a wide range of tools and platforms that facilitate the dissemination of knowledge in novel and unparalleled ways. The prevalence of platforms such as YouTube and Medium may be attributed not just to their accessibility but also to their natural flexibility in accommodating a wide range of content formats, including animated visuals and long-form narratives.

---

*

*E-mail addresses:*

ishan.bangroo@ieee.org (IS Bangroo).

samiaamir.kmm@gmail.com (S Amir).

The field of quantum technology, despite its inherent complexity, offers a unique prospect. Like how art could encapsulate and communicate feelings that are sometimes difficult to express in words, the visual depiction of quantum notions has the potential to bridge the gap between abstract ideas and intuitive understanding. Fundamentally, education may be seen as a process of translation, whereby the intricate language of the cosmos is deciphered and conveyed in a manner that is comprehensible to the human intellect.

In the present environment, visual tools function as a potent means of translation. Consider, as an illustrative example, the concept of Schrödinger's cat, a thought experiment that exhibits a seemingly contradictory nature when described verbally. However, the use of visual representations provides a level of clarity to this perplexing quantum puzzle that may beyond the limitations of verbal communication. Observing the dynamics of shadows inside Plato's allegorical cave is comparable; concepts that first seem abstract and remote soon manifest as palpable realities.

Moreover, narrative frameworks have a twofold function. First and foremost, the authors provide a contextual framework for the material. The expansive domain of quantum science, which is characterized by several phenomena such as superposition, entanglement, and quantum teleportation, may seem fragmented and lacking in interconnectedness. These principles are interconnected via a narrative structure, forming a cohesive fabric where each quantum anomaly serves as a chapter within a broader plot. Furthermore, narratives elicit emotional responses. These sources do not alone provide information, but also serve as sources of inspiration. By contextualizing quantum science within a narrative framework, our objective extends beyond mere understanding to include a sense of awe and fascination, derived from the indescribable excitement that arises when the cosmos unveils one of its enigmatic truths.

The incorporation of these strategies is not only a pedagogical decision, but rather a need in the context of the digital era. In the current day, individuals engaged in the process of acquiring knowledge are faced with an overwhelming abundance of information. Consequently, these learners actively pursue material that could establish a deep connection – content that not only appeals to their intellectual faculties but also evokes an emotional response. As individuals in the field of education, it is our responsibility to adapt and use the extensive array of digital resources available to us, in order to redefine the pedagogy around the teaching of quantum science.

The aesthetic appeal of "QUAVER" resides in its effective combination of these approaches. The integration of visual and verbal representation enhances understanding, and when combined in a cohesive narrative framework, it offers contextualization that makes abstract concepts more real. The counter-intuitive character of quantum events may sometimes be seen as unfamiliar or foreign by many individuals. With visual narratives, our objective is to imbue quantum computing with a sense of humanity, so transcending its purely mathematical and theoretical nature, and instead presenting it as a dynamic and interwoven fabric of narratives.

## 2. Literature review

The idea of engaging students in teaching settings using visual data and storytelling has garnered significant interest. The following are seminal research in closely related fields:

### 2.1. Role of Visualization in Learning from Computer-Based Images

Ainsworth et al. [1] conducted an extensive investigation on the topic of multiple external representations (MERs) in educational contexts, which has contributed to a more profound understanding of the influence of computer-based visualization on the process of learning. The author presents the "DeFT" framework, which stands for Design, Functions, and Tasks, as a means to comprehensively analyze the many ways in which MERs might be used to enhance the educational experience. The research highlights the many functions that MERs may provide, such as presenting several viewpoints on a same topic, juxtaposing complementing material, and facilitating learners' deeper understanding, reconstruction, and connection to the subject matter.

Ainsworth astutely examines the dichotomies that are posed by MERs. On one side, many representations may greatly enhance the understanding of abstract ideas by offering diverse visual aids. However, on the other hand, learners may unintentionally encounter difficulties in effectively integrating these various representations, thereby potentially complicating the learning process. These discoveries have special significance in our research as we start our investigation into the levels of engagement related to information that is enhanced with visual data and narratives.

Ainsworth addresses the educational consequences of her research results. The author proposes that educators and content providers should exercise careful judgment when using visual representations, taking into consideration the appropriateness of including many pictures or diagrams in relation to the desired learning objectives and the cognitive skills of the learners. In the context of our study, this implies a need for systematic and evidence-based methodologies in content creation. It highlights the need of not just including visual data and stories, but also using them thoughtfully to promote meaningful engagement and enhance understanding.

Ainsworth's research explores the cognitive processes that occur when individuals engage with various representations, highlighting the significance of coherence and consistency in visual tales. The author argues that while individual representations may provide limited insights, the integration of these insights across several representations frequently results in a more thorough understanding. The viewpoint is based on the cognitive flexibility hypothesis, which posits that individuals who possess the ability to analyze information from several perspectives and using different methods are more adept at applying their knowledge in a wide range of situations.

Ainsworth asserts the significance of contextualization in relation to the presentation of MERs, emphasizing that the environment in which they are provided is crucial in determining their effectiveness. The comprehension of information may be influenced by several factors such as the setting in which it is presented, whether it is in a digital or physical form, the way representations are sequentially or simultaneously presented, and the amount of engagement afforded with these visual elements. Our study aligns with this viewpoint, as we want to investigate the ideal circumstances for presenting visual data and narratives in order to fully enhance engagement.

A noteworthy aspect of Ainsworth's research is in her investigation of the possible drawbacks associated with over dependence on MERs. The author advises against if an increased number of representations would always result in a higher level of comprehension. In the absence of meticulous design and alignment with learners' preexisting knowledge, the use of many images has the potential to induce cognitive overload, hence impeding the efficacy and efficiency of the learning process. This warning statement is especially relevant to our research, emphasizing the need of finding a middle ground between the complexity of visual materials and the cognitive capacities of our intended recipients.

Ainsworth examinations suggested the implications of her research results for professionals in the field of instructional design and education. The author highlights the need of adopting a strategic methodology that integrates empirical research on learning with principles of design. The dynamic relationship between theory and application is a fundamental tenet that informs our research endeavors. We are committed to ensuring that our findings possess both theoretical relevance and practical usefulness within real-world educational contexts.

*2.2. Narrative as a Learning Tool in Science Centers*

Dahlstrom et al. [2] proposed that there exists a fundamental human predisposition towards narratives, which plays a crucial role in both the retention and comprehension of complicated scientific topics. The research conducted by the author suggests that narratives, due to their inherent structure, provide a framework that facilitates the organization and comprehension of complex material. Through the incorporation of information into familiar circumstances, narratives can render abstract scientific ideas concrete and easily comprehensible, particularly for those lacking previous experience in the given subject matter.

In his analysis, Dahlstrom examines the mechanics of storytelling and highlights the role played by various aspects such as characters, conflicts, resolutions, and emotional arcs. These factors facilitate a cognitive and emotional connection between the information being conveyed and the recipient. These components not only attract attention but also enhance the cognitive processing of the underlying scientific material, making it more accessible and less daunting.

According to Dahlstrom, narratives possess a level of efficacy that beyond ordinary cognition. The findings of his study demonstrate that the use of storytelling has the potential to cultivate a feeling of connection and significance among individuals, hence serving as a catalyst for increased engagement, curiosity, and potential behavioral responses towards the gained information. The presence of emotive elements in narratives plays a crucial role in the field of scientific communication, since its purpose extends beyond the dissemination of information to include the stimulation of curiosity and practical application.

Nevertheless, Dahlstrom also highlights several limitations related to the practice of narrative-based scientific communication. The author places emphasis on the notion that narratives, while serving the purpose of simplification, should refrain from oversimplifying to the extent where misrepresentation occurs. The delicate equilibrium between captivating the audience and maintaining scientific precision is of utmost importance, underscoring the need for scrupulous construction of tales in order to preserve their educational authenticity.

Within the larger framework of our study, Dahlstrom's research offers a fundamental comprehension of the reasons for the substantial improvement in engagement that may be achieved by integrating narratives into material. This statement serves as a reminder that in addition to the visual data, the presentation of information, including its structure, context, and human elements, may significantly impact how it is received and understood by the audience.

Dahlstrom's investigation on the use of story in scientific communication has significant ramifications for the development and dissemination of educational material across many mediums. In a contemporary era characterized by an abundance of information and short attention spans, the skill of integrating scientific facts into captivating narratives may distinguish powerful educational encounters from banal exchanges of information. Narratives serve the dual purpose of appealing to our cognitive inclination towards storytelling and offering a coherent framework that aids in the systematic assimilation of knowledge.

Dahlstrom asserts that narratives possess a universal quality. Scientific ideas may be effectively communicated via the use of visual representations, such as diagrams and illustrations, which can transcend many cultural, language, and educational boundaries. This inclusive nature of visual mediums enables the communication of complicated scientific concepts to a wide range of individuals. When people see their own experiences within a narrative framework, they not only enhance their comprehension of scientific concepts, but also develop a deeper appreciation for the applicability of such knowledge to their own personal circumstances. The establishment of an emotional connection has the potential to stimulate further investigation, analytical reasoning, and a heightened level of admiration for the topic at hand.

Given the dynamic nature of the educational environment, the incorporation of narratives has the potential to greatly enrich virtual and augmented reality encounters, interactive displays, and e-learning modules. The comprehension of the intricacies of storytelling and its mutually beneficial connection with scientific communication, as emphasized by Dahlstrom, offers a framework for developing experiences that have a profound impact, facilitate learning, and foster motivation.

Dahlstrom's focus on narrative-driven techniques serves to reinforce the viability of including storytelling components into our research methodology, aligning with our own investigations. As the investigation progresses, it becomes apparent that the integration of data and story frameworks has the potential to significantly transform the way audiences perceive, engage with, and retain scientific information.

*2.3. Influence of Visualization on Abstract Concept Comprehension*

Mayer's comprehensive exploration of the cognitive processes that underlie multimedia learning establishes a significant milestone in understanding the complex interplay between textual information and visual elements within educational contexts. The foundation of his investigation [3] is in the dual-coding theory, a theoretical framework that suggests the distinct and simultaneous processing of visual and verbal information in human cognition. The significance of this differentiation is in its ability to enable multimedia presentations to effectively accommodate the needs of both visual and auditory learners concurrently, hence enhancing the overall learning experience.

Mayer provides a comprehensive analysis of the ways in which images serve to complement and enrich textual content.

For example, diagrams can provide spatial representations of abstract ideas, so enhancing their tangibility and relatability. In addition, animations have the capability to visually depict dynamic processes occurring over a period, so facilitating learners in comprehending the sequential or cyclical characteristics inherent in certain phenomena. The integration of words and visuals in an educational setting promotes a heightened level of immersion and engagement, resulting in enhanced retention and comprehension.

Mayer's observations also address the possible challenges associated with multimedia learning. The author advises against cognitive overload, highlighting the potential negative effects of using pictures without discretion. Instead of facilitating understanding, an excessive use of visuals might overwhelm and perplex learners. Hence, it is essential to maintain a prudent equilibrium, guaranteeing that visual elements fulfill a specific objective and are congruent with the written material.

The results of Mayer have substantial ramifications within the current educational environment that is heavily influenced by digital technology. With the continuous advancement and enhanced availability of multimedia technologies, educators and content producers now have a wide range of formats and platforms at their disposal. Nevertheless, it is crucial to comprehend the cognitive underpinnings, as delineated by Mayer, in order to use these tools in an efficient manner.

When considering the relevance of Mayer's findings to our own study, it becomes apparent that our focus on visualization should be supported by a comprehensive comprehension of cognitive mechanisms. In the process of developing our methodology and tools, Mayer's insights serve as a valuable reference, emphasizing the significance of well-constructed multimedia material. This statement emphasizes the need of considering both the content and the way visual elements interact with textual information in order to optimize the advantages they provide, resulting in a coherent, intuitive, and fulfilling educational encounter.

Expanding upon the research conducted by Mayer, it is crucial to acknowledge the dynamic characteristics of multimedia platforms and the growing heterogeneity among learners. In the era of digital transformation, multimedia is not just seen as a tool, but rather as an interconnected ecosystem in which many modalities live in harmony. The use of interactive simulations, virtual reality, and augmented reality is revolutionizing the parameters of educational encounters, exposing learners to immersive settings that blend the distinctions between tangible reality and virtual realms.

The improvements are significantly transforming the practical use of Mayer's dual-coding theory in the field. For example, Mayer's research primarily emphasized the equilibrium between textual information and still images. However, in the present period, there is a growing emphasis on the use of dynamic visuals, immersive three-dimensional (3D) settings, and tactile encounters to enhance the process of learning. These improvements provide a potential avenue for catering to the varied learning requirements of people who may not neatly fit into the categories of 'visual' or 'auditory' learners, but may derive advantages from engaging in multisensory experiences.

Nevertheless, the pursuit of these prospects is not without its share of hurdles. The potential for cognitive overload is not just confined to the interplay between textual content and visual elements, but also encompasses the intricacy of the multimedia technologies used. The extensive level of engagement offered by contemporary educational technologies, however captivating, may sometimes undermine the fundamental educational goals if not well crafted. In this context, Mayer's admonitions against cognitive overload have heightened significance. Educators must possess the ability to effectively evaluate and determine the suitability of various multimedia technologies in accordance with their individual instructional objectives and the unique requirements of their pupils.

The increasing prevalence of collaborative and socially interactive multimedia learning environments necessitates a comprehensive understanding of the influence of group dynamics and social signals on individual learning processes within these settings. The engagement of individuals with material is not the exclusive focus, but their interaction with peers inside these contexts is equally significant.

Mayer's study provides a fundamental framework for examining multimedia learning, although it is incumbent upon contemporary educators and researchers, such as ourselves, to consistently modify and advance these principles. By establishing a solid foundation in robust cognitive theories, the integration of advanced multimedia technologies may facilitate the development of enhanced educational experiences that are more efficacious, inclusive, and transformational.

*2.4. Leveraging Multimedia in Higher Education*

By examining the influential research conducted by Clark and Mayer [4], one may get a deeper understanding of the intricate process of curating instructional information in the context of the digital era. The advent of several e-learning platforms has brought about a significant transformation in the higher education sector, leading to a move from conventional lecture halls to online classrooms that transcend geographical limitations. The principles provided by the pair provide a significant framework for instructors, course designers, and institutions interested in using multimedia to improve the learning process.

The book thoroughly emphasizes the need of adopting a well-balanced strategy, while also warning against the excessive use of attention-grabbing animations or unnecessary images that may hinder rather than facilitate the learning process. On the contrary, they advocate for the intentional incorporation of multimedia components, prioritizing excellence rather than abundance. For example, animations have the potential to render abstract ideas more physical. However, it is crucial that these animations maintain a strong alignment with the intended learning goals and are presented at suitable points in the instructional process.

Moreover, Clark and Mayer extensively examine the complexities of cognitive processing, emphasizing the manner in which learners engage with multimedia materials. The authors illuminate the significance of minimizing superfluous cognitive burden, underscoring the notion that multimedia should aim to streamline rather than complicate the process of learning. This assertion has significant relevance in the current age of excessive information, when learners are consistently exposed to a wide range of stimuli, hence posing difficulties in maintaining concentration throughout the learning process.

The emphasis placed on narratives by the subject under consideration offers a compelling route for further investigation. Narratives have long been recognized as potent instruments of communication, and their use into e-learning modules may provide a unifying element, enhancing the memorability and relatability of the learning experience. By incorporating real-world situations and case studies into the curriculum, learners can more effectively situate theoretical ideas within practical contexts, resulting in enhanced comprehension and long-term recall.

Furthermore, the book explores the changing dynamics of e-learning, including the emergence of mobile learning, gamification, and adaptive learning paths. These ideas have great significance in a contemporary society where education is progressively adapting to individualized approaches, accommodating diverse learning speeds, preferences, and prior knowledge.

The guide authored by Clark and Mayer serves not only as a theoretical discourse, but also as a practical manual for those involved in the realm of higher education. As scholars engage with the diverse landscape of multimedia within academic contexts, their research serves as a guiding light, ensuring that we stay grounded in effective instructional approaches. The work presented by the authors encourages individuals to engage in a combination of creative and analytical thinking. This approach aims to create learning settings that are both immersive and cognitively sensitive, as well as dynamic and goal-oriented, reflecting the modern demands of the 21st-century learner.

The integration of technology into the field of education has not only broadened the scope of information sharing, but has also altered instructional methods. This transformation is discussed by Clark and Mayer with a feeling of urgency and meticulousness. The investigation into the domain of e-learning, instead of being prescriptive, presents a structured strategy that assists educators in navigating the many obstacles and possibilities associated with digital education.

One of the significant observations inherent in their work is the interplay and harmonious relationship among content, technology, and the human factor. Although technological tools and platforms continue to advance at a fast pace, the fundamental structure of human cognitive architecture remains essentially unchanged. Educators are compelled to find a balance between using advanced multimedia technologies and acknowledging the inherent cognitive abilities and constraints of learners. This statement serves as a reminder that despite the increasing sophistication of our instruments, the fundamental objective remains unaltered: to allow authentic comprehension and long-lasting knowledge.

The focus placed by Clark and Mayer on interaction highlights the shift from passive learning to active participation. Within the world of digital platforms, individuals engaged in the process of learning are not only passive recipients of information, but active participants in the generation and shaping of knowledge. The book serves as a source of inspiration for educators, encouraging them to develop instructional approaches that cultivate critical thinking, problem-solving skills, and active engagement. This shift in pedagogy involves moving away from the traditional method of rote memorization and embracing a more exploratory and inquiry-based framework.

Within the wider framework of higher education, their research and contributions stand as a monument to the dynamic and ever-changing nature of the academic sphere. The dissolution of traditional boundaries is evident, as there is a notable movement towards the importance of lifelong learning, multidisciplinary investigation, and collaborative knowledge generation. According to the authors, it is the responsibility of educators and institutions to ensure that multimedia is used in a prudent manner, with the aim of enhancing the learning process rather than diminishing its effectiveness.

The ideas presented by Clark and Mayer extend beyond a mere plan, as they give a visionary perspective on the potential aspirations of modern education. By integrating the principles of cognitive science, technology, and pedagogy, the authors construct a comprehensive framework that offers a multitude of potential outcomes. Their work encourages us to reconsider the fundamental nature of education and the process of acquiring knowledge in the era of digital advancements. The work presented serves as a compelling appeal to both educators and instructional designers, advocating for an education system that is captivating, efficient, and continuously progressing.

The integration of visualization, narrative strategies, cognitive theories, and multimedia components within educational frameworks has garnered growing attention in contemporary pedagogical research. The importance of working examples in facilitating students' understanding of complex topics has been underscored by scholars such as Anderson and Schunn [5]. Moreover, Schwartz and Arena (2013) expounded on the need for a harmonious integration of discovery-based learning and direct teaching, emphasizing the significance of visual aids in both domains. The influence of narration in the field of education is not a recent discovery. In fact, the study conducted by Bruner [7] has long emphasized the importance of narrative in the process of cognitive development. In addition to these methodologies, the use of multimedia has been advocated by scholars such as Paivio[8] as a means to promote active involvement and long-term retention. According to Paivio's theory of dual coding, the combination of visual and aural stimuli synergistically promotes memory formation and comprehension. Collectively, these studies provide a comprehensive perspective on the educational environment, confirming the need for instructors and designers of educational materials to integrate visuals, narratives, and multimedia elements in a synergistic manner.

## 3. Methodology

### 3.1. Objectives

The primary aim of this study is to conduct a thorough assessment of the efficacy of using visualization methods, narrative structures, and multimedia components as instructional aids to enhance students' understanding and level of involvement in higher education environments.

In order to accomplish this objective, we suggest doing an in-depth exploration of several facets associated with the improvement of learning outcomes. These dimensions include a wide range of factors, which may include, but are not limited to:

The objective of this study is to investigate if the inclusion of these aspects may enhance learners' memory retention compared to conventional instructional approaches.

*Involvement Levels*: This study aims to assess the degree of student involvement when exposed to learning materials that include visual, narrative, and multimedia aspects. The objective is to determine if these tools have the potential to cultivate a more captivating and dynamic learning environment.

*Critical Thinking*: The objective is to evaluate if the integration of various techniques facilitates the development of critical thinking skills by enhancing students' ability to analyze and synthesize information in a more efficient manner.

*Applicability in Diverse Disciplines*: This study aims to assess the adaptability of these methodologies in various academic fields, determining their efficacy in both scientific disciplines and the humanities or arts.

*Teacher's Perspective*: In order to get knowledge from educators on the possible advantages and difficulties of incorporating these tactics into the curriculum, and to examine their viewpoints on how these tools might be maximized to improve educational results.

The purpose at hand is firmly rooted in the theoretical frameworks that have been emphasized in the literature review. It aims to expand upon the existing fundamental knowledge by presenting facts and insights that have been collected from educational settings in the real world. The primary objective of this study is to provide a meaningful contribution to the current pool of knowledge and provide practical recommendations for educators and policymakers who are interested in improving the educational experience by using visualization, story, and multimedia technologies.

This research aims to cultivate an enhanced and inclusive educational setting that accommodates the many learning preferences and requirements of contemporary students. Consequently, it seeks to enable the development of more sophisticated and efficient teaching methods for the future of higher education.

*3.2. Research Design*

*Mixed-Methods Approach:*
Considering the complex nature of our goal and the wide range of literature highlighting the influence of visualization, story, and multimedia on the learning process, we have opted for a mixed-methods study approach. This methodology integrates quantitative surveys and qualitative focus groups in order to provide a thorough and holistic understanding of the phenomena.

- Quantitative Surveys:
The purpose of this investigation is to collect empirical data that quantifies the efficacy of using visualization, story, and multimedia components in improving students' understanding and level of involvement in the learning process.

An assortment of students from various academic fields in tertiary education. Given the wide range of literature including Ainsworth's visualization framework and Clark and Mayer's guidance on multimedia, the survey sample will be composed of students from both the sciences and humanities, thereby assuring a comprehensive and equitable representation.

The research will use structured questionnaires that will be developed using Likert scale questions. These questionnaires will aim to assess the participants' levels of comprehension, retention, involvement, and overall satisfaction with the learning experience. The design will be informed by Mayer's cognitive theory of multimedia learning in order to correctly capture the effects of visuals and multimedia.

- Qualitative Focus Groups:
The objective of this study is to conduct a comprehensive exploration of the firsthand encounters of students and educators while engaging with various educational resources. The primary objective of our study is to use open-ended talks as a means of uncovering the subtleties, difficulties, and potential advantages that may not be readily apparent when relying just on quantitative data.

A wide range of students, educators, and instructional designers will be selected in order to include a diverse array of views. The focus on narratives as emphasized by Dahlstrom will be specifically examined within the context of humanities students, while Mayer's thoughts on multimedia will be investigated primarily among scientific majors and educators.

The tools used in this study will consist of carefully curated open-ended discussion guides that focus on the experiences, perspectives, and suggestions of the students. From the perspective of educators, the focus of talks will be on the deployment of these tools, the level of student involvement seen, and the possible pedagogical changes that may arise as a result of their integration.

The use of a dual strategy facilitates the augmentation of the available data reservoir. The use of quantitative data allows for the measurement of metrics pertaining to the effectiveness of tools. On the other hand, qualitative insights provide a more comprehensive comprehension of the experiences and preferences of both students and instructors. This combination of approaches provides a complete depiction of the existing state and possible future developments in this field.

*3.3. Sample*
In accordance with our aim to evaluate the impact of integrating visualization, story, and multimedia technologies on understanding, it is essential to obtain a sample that sufficiently encompasses the variety seen within the realm of higher education. Consequently, three educational establishments, exhibiting diversity in terms of their size, geographical placement, and academic emphasis, have been chosen.

- Criterion:

➢ Size and Scale: Institutions spanning from large universities with diverse faculties to smaller colleges with a focus on specific disciplines were evaluated to determine the efficacy of our methods across varying operational scales.

➢ Geographic Diversity: Considering institutions from urban, suburban, and rural contexts in order to comprehend how geographical location and associated socioeconomic factors may affect student engagement with multimedia learning tools.

➢ Curriculum Diversity: Because our literature ranges from Dahlstrom's narrative-based insights to Mayer's multimedia learning principles, institutions that offer a wide range of courses, from the sciences to the arts, were chosen.

- Stratified Random Sampling:

  ➢ The purpose of using stratified random sampling is to enhance the representativeness of our sample, so ensuring that it accurately reflects the broader population of students within the realm of higher education. This facilitates the attainment of conclusions that are both precise and applicable to a wider range of contexts.

  ➢ In the selected institutions, the various departments or faculties will be classified as distinct strata. A stratified random sampling technique will be used to choose a representative sample of students, maintaining inclusivity across different academic fields, year groups, and demographic backgrounds.

  ➢ One advantage of using this approach is that it helps to mitigate the potential bias introduced by certain demographic groups, hence enhancing the robustness and reliability of our research results. The proposed study would include a range of diverse topics, including Ainsworth's visualization approaches and Clark and Mayer's multimedia methodologies.

### 3.4. Data Collectivity:

- *Survey Methodology:*
  Based on the previously established significant digital engagement measures, a systematic set of surveys was implemented. The objective of these surveys was to:

    ▪ Quantitative Assessment of User Engagement: Correlate the input received from students and instructors with the tangible online indicators that were monitored, such as LinkedIn impressions, Medium.com subscribers, and website visits.
    ▪ Assessed Effectiveness: Assess the degree to which participants assessed their comprehension and retention of material while employing integrated learning tools in comparison to conventional approaches.
    ▪ Gathering Qualitative Feedback: Integrated Learning Experience: Gain an understanding of user experiences related to the integration of imagery, narrative, and multimedia elements in educational materials.
    ▪ Request for Recommendations: Seek suggestions and enhancements for optimizing the current system to effectively cater to the needs of both learners and instructors.

In order to enhance our technique, we utilized scatter plot (*Fig.1*) to graphically depict the correlation between content quality, engagement levels, and the integration of visual data and narratives in the material. The horizontal axis will reflect the quality of material, while the vertical axis will show the degrees of engagement (in percentage). The points on the graph will be colored using gradients to illustrate the breadth of visual data and stories included, with warmer colors indicating larger quantities.

The scatter plot demonstrates a favorable link between the presence of high-quality information, including visual data and stories, and enhanced engagement. This finding supports our initial premise.

The centroids of the clusters, denoted by the symbol 'X', represent the mean values of engagement and content quality across different degrees of visual data and narratives used. The conformity of these centroids to the indicated trend provides more support for our theory.

The proposed technique is to provide a comprehensive understanding of how the combination of visualization, story, and multimedia technologies might reinvent learning experiences in higher education. This will be achieved via the amalgamation of statistical analysis, personal experiences, and qualitative insights.

### 3.5. Engagement Analysis Model (EAM):

In order to get a comprehensive understanding of the complex interplay of engagement levels, content quality, and the incorporation of visual data and stories, we propose the introduction of the Engagement Analysis Model (EAM). The foundation of this approach is on re-iterative DL algorithms, and its primary objective is to accurately predict engagement metrics. The algorithmic technique may be succinctly expressed as follows:

*AIM:*
The core objective of the EAM (Engagement Analysis Model) is to forecast levels of engagement via the examination of variables related to content quality, as well as the inclusion or exclusion of visual data and narratives.

*Data Preparation:*
The dataset is partitioned into two subsets: a training set, denoted as "$D_{train}$" and an evaluation set, denoted as "$D_{eval}$" The training set is used to facilitate the training process of the model and progressively enhance its parameters. On the other hand, the assessment set is used to assess the model's performance in terms of accuracy and the F1 score.

*Optimization Cycle:*
The process of training and evaluating is performed iteratively, with a predetermined number of epochs ($E$). The iterative nature of this cycle guarantees that the model consistently acquires knowledge and improves its predictive capabilities, hence maximizing its overall performance as time progresses.

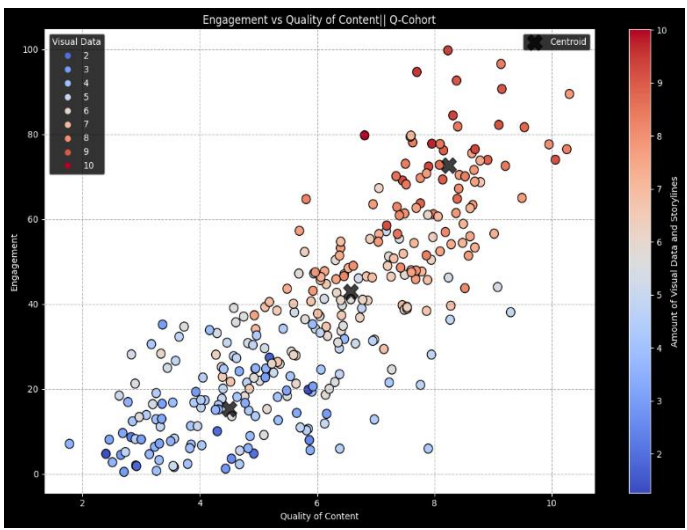

**Fig. 1.** Graphical Representation of Quality of content vs Engagement

**Engagement Analysis Model (EAM) Algorithm:**
1. **Input:**
   - Training dataset $D_{train}$ (containing engagement metrics, content quality scores, and indicators of visual data/storylines inclusion).
   - Evaluation dataset $D_{eval}$.
   - Number of epochs E.
   - Learning rate LR.
   - Optimizer OPT.
   - Fine-tuning parameters $\theta_{ft}$.
2. **Initialize the EAM model** with parameters $\theta$.
3. **For each epoch e from 1 to E**, execute the following:
   a) **Training & Fine-tuning:**
   i. For every batch $B_{train}$ in $D_{train}$:
   - Compute the EAM model's predictions based on the input batch $B_{train}$.
   - Calculate the loss L between the model's predictions and actual engagement scores.
   - Update the model's parameters $\theta$ using optimizer OPT and learning rate LR.
   ii. Implement fine-tuning on the EAM model with parameters $\theta_{ft}$.
   b) **Evaluation:**
   i. Begin with performance metrics: accuracy Acc and F1.
   ii. For every batch $B_{eval}$ in $D_{eval}$:
   - Compute the EAM model's predictions based on the input batch $B_{eval}$.
   - Determine the accuracy and F1 score using the model's predictions and actual engagement scores.
   - Update and aggregate the overall metrics Acc and F1.
   iii. Output: Overall accuracy Acc and F1 score for epoch e.
4. **Output:**
   - Return the model parameters $\theta$ that produce the highest evaluation accuracy Acc over all epochs.
   - Provide comprehensive reports on patterns identified, highlighting relationships between content quality, visual data/storyline inclusion, and engagement levels.

*OUTCOME:*
At the termination of each epoch, the model's parameters that yielded the greatest accuracy are preserved. The optimized parameters represent the model's most accurate interpretation of the correlation between content quality, presence of visual data/storyline, and degrees of engagement.

4. **Resulting amalgam of Anticipated Outcomes**

   The objective of our comprehensive technique is to get a comprehensive comprehension of engagement metrics pertaining to instructional material that is enhanced with visualizations, narratives, and multimedia components.

   *VALIDATING HYPOTHESIS*

   The hypothesis is central to the core of this investigation. Based on the first observations and statistical studies, it can be inferred that there exists a significant relationship between the quality of material, the incorporation of visual data and stories, and the level of engagement it generates. This inference is supported by the rejection of the null hypothesis in favor of the alternative hypothesis. By rejecting the null hypothesis, it has been statistically shown that there exists a large disparity in levels of engagement when information is enhanced with visual data and stories.

*INSIGHTS:*

The Engagement Analysis Model (EAM), supported by our AI-driven methodology, is meant to further demonstrate the value of combining these three forms of content delivery. The predicted results of the model will shed light on:

- How much text and how much graphic material works best for various student groups.
- Ways to improve content that have a noticeable effect on reader participation.
- Teaching methods that improve students' understanding and memory.

*Actionable Recommendations:*

Despite statistical validation, our technique is designed to provide practical discoveries that have the potential to revolutionize instructional practices. Educational institutions, as well as content developers and ed-tech platforms, can use these valuable insights in order to customize their contents and teaching approaches. This will guarantee that learners are not just actively involved but fully immersed in a comprehensive learning experience that is meaningful and enduring.

5. **Conclusion**

Our research is situated at the intersection of scientific rigor and educational improvement, as the field of education continues to expand with the integration of technology and the implementation of novel pedagogical techniques. By using a systematic methodology, our objective extends beyond mere statistical validation, as we want to instigate a transformative change in the perception, generation, and consumption of educational materials. The prospective trajectory of education entails a paradigm shift towards an interactive, engaging, and profoundly effective approach. Through this study, our aim is to provide valuable guidance and practical approaches for the international educational community by utilizing data-driven analysis and strategic recommendations.

**ACKNOWLEDGEMENT**

We wish to convey our profound gratitude to Womanium. Our research was enriched immeasurably through our alliance as Team Q-Cohorté, gleaning pivotal data and invaluable insights from this association.

The tireless endeavors of the Womanium in promoting inclusion within STEM education and their everlasting commitment to equalizing opportunities for everyone. The alignment between their objective and ours not only easily integrates but also enhances the resonance of our results, hence increasing the relevance within the current educational context.

The principles of inclusiveness, empowerment, and forward-thinking embodied by Womanium have significantly impacted the trajectory of our study. Being associated with WOMANIUM as Q-COHORTÉ, our involvement in this significant worldwide endeavor has served as a guiding principle in our efforts to improve pedagogy in the field of quantum education.

We express our utmost gratitude to the whole team at the Womanium for their consistent support, advice, and collaborative efforts. Collectively, we aspire that our collaborative endeavors will facilitate the progression towards an educational future that is characterized by inclusivity, engagement, and efficacy for all individuals.

Within the contemporary academic landscape, characterized by the rapid progressions in technology and quantum learning, it is not uncommon to find oneself enthralled by the newest terms, much like a youngster entranced by a novel and gleaming plaything. Nevertheless, it is important to acknowledge that despite the fascination with artificial intelligence and technical terminology, the fundamental importance lies in the genuineness and ethical nature of our methodology. In the expansive realm of information, wherein artificial intelligence (AI) and other sophisticated procedures assume a crucial role, it is the authentic commitment to study, the meticulous application of rigorous methodology, and the sincere pursuit of knowledge that really engender significant transformations. Our article serves as a manifestation of this conviction.

Efforts have been made to ensure that our research results possess not just statistical significance but also a firm grounding in the practical ramifications, aiming to shape a future in which education is inclusive, engaging, and successful for all individuals. It is expected that our readers would recognize the sincerity of our devotion and see our thoughts as a valuable contribution and a compelling call to action within the realm of quantum education.